\documentclass[11pt]{article}
\usepackage{amsfonts}
\usepackage{graphicx}
\usepackage{epstopdf}
\usepackage{epsfig}
\usepackage{amssymb}
\usepackage{setspace}
\usepackage{caption}
\usepackage{amsfonts}
\usepackage{color}
\usepackage{amsmath}
\usepackage{float}
\usepackage{comment}
\newcommand {\be}{\begin{equation}}
	\newcommand {\ee}{\end{equation}}
\newcommand {\bea}{\begin{array}}
	
	\newcommand {\eea}{\end{array}}
\newcommand{\RN}{Reissner-Nordstrom~}

\textwidth=6.5in \hoffset=-.75in \voffset=-1in \numberwithin{equation}{section}
\numberwithin{figure}{section}

\begin{document}

	\begin{titlepage}
			\vspace{1cm} 
			\begin{center}
				{\Large \bf {Hidden conformal symmetry and pair production near the cosmological horizon in Kerr-Newman-Taub-NUT-de Sitter spacetime}}\\
			\end{center}
			\vspace{2cm}
			\begin{center}
				\renewcommand{\thefootnote}{\fnsymbol{footnote}}
				Haryanto M. Siahaan{\footnote{haryanto.siahaan@unpar.ac.id}}\\
				Center for Theoretical Physics, Department of Physics,\\
				Parahyangan Catholic University,\\
				Jalan Ciumbuleuit 94, Bandung 40141, Indonesia
				\renewcommand{\thefootnote}{\arabic{footnote}}
			\end{center}
			\vspace{2cm}
			\begin{abstract}
				
We show that the study of hidden conformal symmetry that associates to the Kerr/CFT correspondence can also apply to the cosmological horizon in Kerr-Newman-Taub-NUT-de Sitter spacetime. This symmetry allows one to employ some two dimensional conformal field theory methods to understand the properties of the cosmological horizon. The entropy can be understood by using the Cardy formula, and the equation for scattering process in the near region has an agreement with the one obtained from a two point function in two dimensional conformal field theory. We also show that pair production can occur near the cosmological horizon in Kerr-Newman-Taub-NUT-de Sitter for near extremal conditions.
				
			\end{abstract}
		\end{titlepage}\onecolumn 
		\bigskip 
		
\section{Introduction}\label{sec.intro}
		
Astronomical observations suggest the expansion of our universe. In a relativistic theory of gravity, the expanding spacetime is described by the de Sitter solution \cite{Peebles:2002gy,Padmanabhan:2002ji} that solves the Einstein equations of motions with a positive cosmological constant \cite{Griffiths:2009dfa}. In the de Sitter spacetime, there exists the so-called cosmological horizon that behaves similar to black hole horizon in many ways. For example, one can compute the Hawking temperature associated to this horizon \cite{Gibbons:1977mu}. Consequently, there also exists a first law of thermodynamics relation for this horizon as proposed in \cite{Teitelboim:2001skl,Gomberoff:2003ea,Sekiwa:2006qj}. Then it is understood that if a de Sitter spacetime contains black hole, then the spacetime is equipped with multiple horizons.

The Kerr/CFT correspondence is used to understand some aspects of black hole horizon \cite{Guica:2008mu,Hartman:2008pb,Compere:2012jk}. The warped AdS structure in the near horizon geometry of an extremal black hole opens the possibility to use some two dimensional conformal field (CFT$_2$) theory methods in understanding several aspects of black hole. For example, the Cardy formula for entropy in a CFT$_2$ can recover the Bekenstein-Hawking entropy of the black hole. Another one is the scattering near horizon which can also be described by using a two point function calculation in a CFT$_2$. The Kerr/CFT correspondence can also be extended to the non-extremal case where the conformal symmetry is hidden in the test scalar wave equation. Related to the Cardy formula in non-extremal setup, we assume that the corresponding central charge does not change as the geometry evolves from the extremal state. In literature, Kerr/CFT has been discussed for many cases of black holes that can exist in various theories of gravity \cite{Compere:2012jk,Ghezelbash:2012qn,Ghezelbash:2014aqa,Sakti:2019zix,Sakti:2020jpo}. 

Various similar properties between the black hole and cosmological horizons motivate us to establish the Kerr/CFT correspondence for the cosmological horizon. In literature, there exists a particular holography for de Sitter spacetime \cite{Strominger:2001pn} that relates quantum gravity on D-dimensional de Sitter space to conformal field theory on a single (D-1)-sphere. However, in this work, we would like to employ the Kerr/CFT correspondence to the cosmological horizon in Kerr-Newman-Taub-NUT-de Sitter spacetime that has several parameters, namely the rotation, electric charge, NUT parameter, and a positive cosmological constant. Here we also anticipate the newly proposed first law of thermodynamics for a NUT spacetime \cite{Wu:2019pzr,Wu:2022rmx} where it introduces some new conserved charges as thermodynamic parameters in the system. Following \cite{Sekiwa:2006qj}, we also consider the cosmological constant that can vary \cite{Caldarelli:1999xj} which modify the first law of cosmological horizon mechanics. 

On the other hand, the pair production can occur near the horizon of near extremal as a Schwinger effect \cite{Chen:2012zn,Siahaan:2019ysk,Chen:2020mqs,Zhang:2020apg,Chen:2021jwy} in analogy to such effect in strong electromagnetic field. It can be shown that the violation of the Breitenlohner-Freedman bound that associates to the massive scalar field near the horizon of the near extremal black hole leads to the conclusion that the pair production can exist. The resemblances between black hole and cosmological horizons suggest we investigate such an effect for the cosmological horizon in Kerr-Newman-Taub-NUT-de Sitter spacetime. To the best of our knowledge, such study has not appeared in literature.

The organization of this paper is as follows. In the next section we provide a short review on Kerr-Newman-Taub-NUT-de Sitter spacetime. The twofold hidden conformal symmetry is constructed in section \ref{s3.hidden}. The microscopic entropy calculation is performed in section \ref{s4.microentropy}, whereas the supporting holographic scattering discussion is worked out in section \ref{s5.scattering}. The study of pair production near the cosmological horizon is performed in section \ref{sec.pair}. Finally, a conclusion is given. In this paper we consider the natural units $G = {\hbar} = k_B = c = 1$. 
		
\section{Kerr-Newman-Taub-NUT-de Sitter spacetime}\label{sec.KNTNdSreview}
		
Kerr-Newman-Taub-NUT-de Sitter (KNTNdS) spacetime solves the Einstein-Maxwell equations of motion with a positive cosmological constant, namely
\be \label{eq.Einstein}
R_{\mu \nu }  - \frac{1}{2}g_{\mu \nu } R + \frac{3}{l^2} g_{\mu \nu }  = 2F_{\mu \alpha } F_\nu ^\alpha   - \frac{1}{2}g_{\mu \nu } F_{\alpha \beta } F^{\alpha \beta } \,,
\ee 
where $F_{\mu\nu}=\partial_\mu A_\nu -\partial_\nu A_\mu $. The vector field obeys the source free condition $\nabla _\mu  F^{\mu \nu }  = 0$ and the Bianchi identity $\nabla _\mu  F_{\alpha \beta }+\nabla _\beta  F_{\mu \alpha }+\nabla _\alpha  F_{\beta \mu }=0$. The KNTNdS spacetime metric can be expressed as 
\be \label{metricKNTNdS}
ds^2  =  - \frac{{\Delta _r }}{{\rho ^2 }}\left( {dt - \left( {a\Delta _x  - 2nx} \right)d\phi } \right)^2  + \rho ^2 \left( {\frac{{dr^2 }}{{\Delta _r }} + \frac{{dx^2 }}{{\Delta _x \Delta _l }}} \right) + \frac{{\Delta _x \Delta _l }}{{\rho ^2 }}\left( {adt - \left( {r^2  + a^2  + n^2 } \right)d\phi } \right)^2
\ee 
where $\Delta_x = 1-x^2$,
\be
\Delta _r  = r^2  - 2Mr + Q^2  + a^2  - n^2  - \frac{3}{{l^2 }}\left( {\left( {a^2  - n^2 } \right)n^2  + \left( {\frac{{a^2 }}{3} + 2n^2 } \right)r^2  + \frac{{r^4 }}{3}} \right)\,,
\ee 
\be 
\Delta _l  = 1 + \frac{ax\left(4n+ax\right)}{l^2} \,,
\ee 
and $\rho ^2  = r^2  + (n+ax)^2 $. It is understood that $M$, $a$, $n$ and $Q$ are the mass, rotation, NUT charge, and electric charge parameters. The corresponding cosmological constant in the equation above is $\Lambda = 3 l^{-2}$. The accompanying vector components to solve the Einstein equations (\ref{eq.Einstein}) are
\be \label{eq.vector}
A_\mu dx^\mu = \frac{Qr}{\rho^2} \left( dt + \left(2n x - a \Delta_x \right)d\phi\right)\,.
\ee 

\begin{figure}
	\begin{center}
		\includegraphics[scale=0.5]{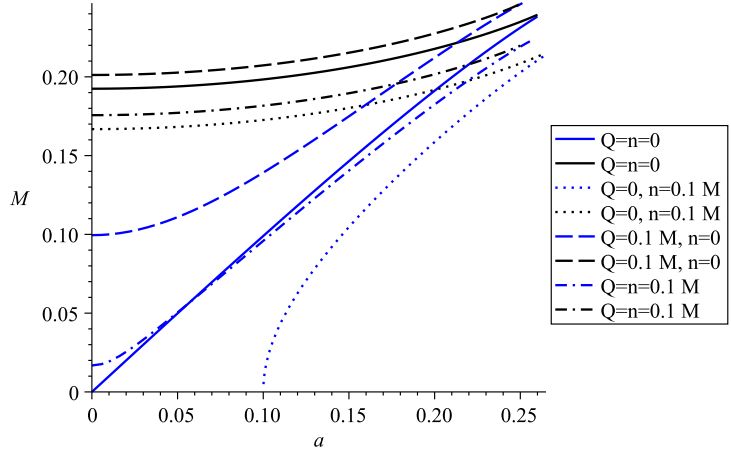}
	\end{center}
	\caption{Case $l=M$. Black curves represent $r_c=r_b$, whereas blue ones for $r_b=r_i$.}\label{fig.extreme}
\end{figure}

\begin{figure}
	\begin{center}
		\includegraphics[scale=0.5]{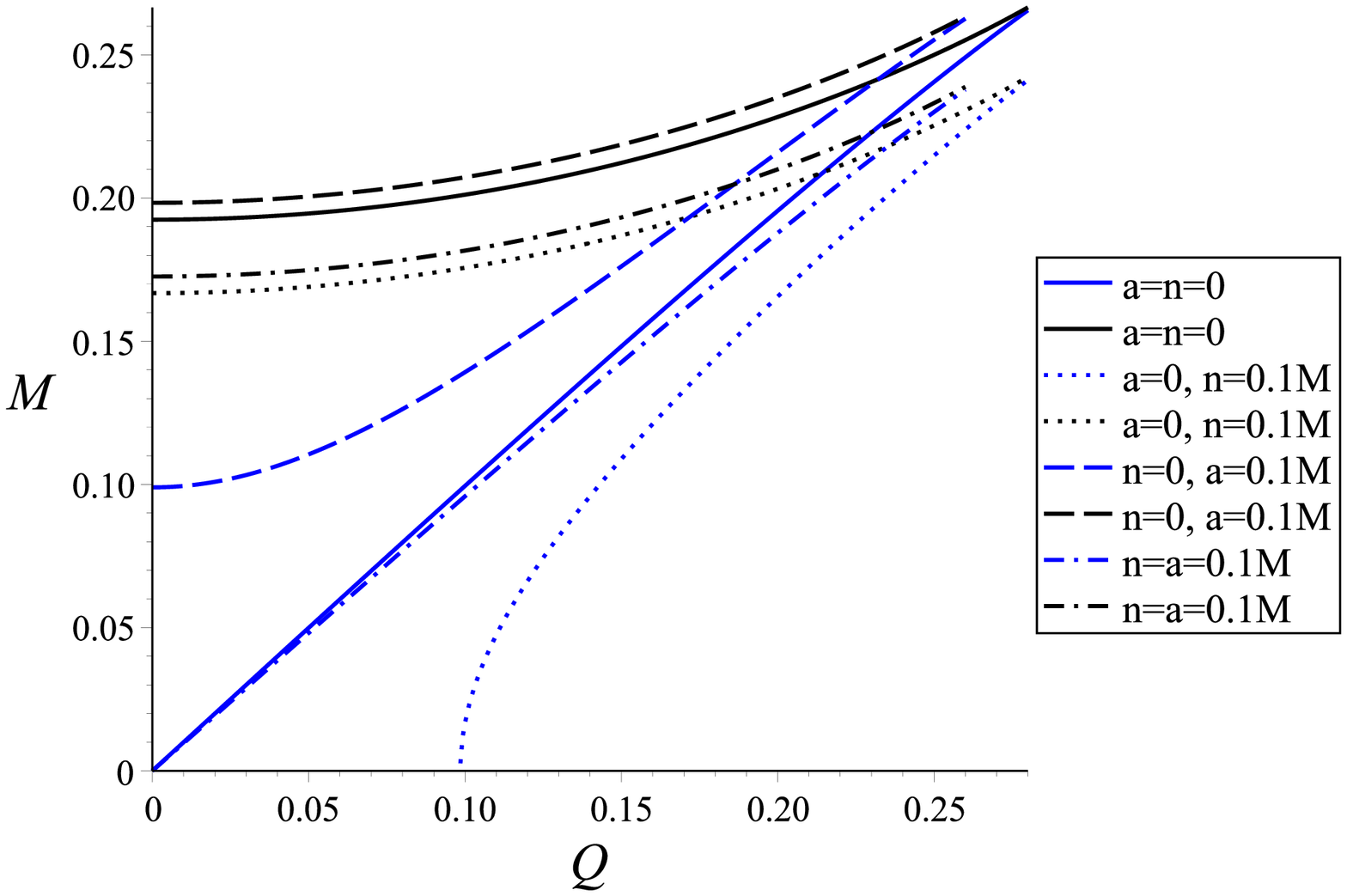}
	\end{center}
	\caption{Case $l=M$. Black curves represent $r_c=r_b$, whereas blue ones for $r_b=r_i$.}\label{fig.extremeQ}
\end{figure}

\begin{figure}
	\begin{center}
		\includegraphics[scale=0.5]{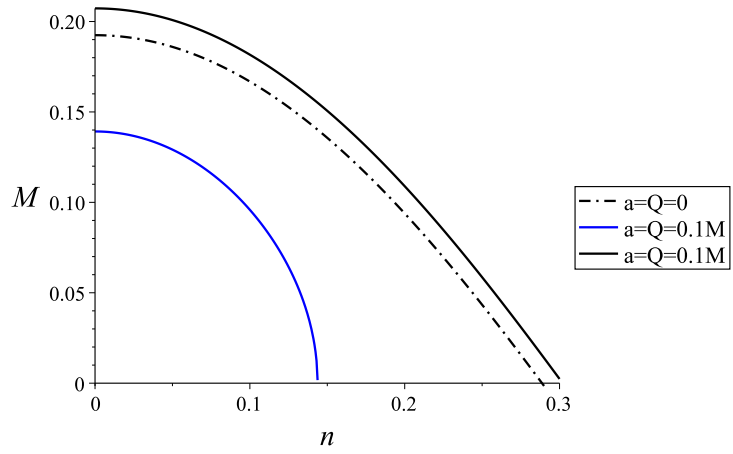}
	\end{center}
	\caption{Case $l=M$. Black curves represent $r_c=r_b$, whereas blue ones for $r_b=r_i$. The dashed blue line does not exist since the case of $a=Q=0$ describes the Taub-NUT-de Sitter spacetime.}\label{fig.extremen}
\end{figure}

It is known that multiple horizons with positive radii can exist in Kerr-Newman-Taub-NUT-de Sitter spacetime. The largest one is acknowledged as the cosmological horizon $r_c$, the outer black hole horizon is $r_b$, and the inner one is denoted by $r_i$, where $r_c > r_b > r_i$. The coincidence of cosmological and outer black hole horizons is known as the extremal state for cosmological horizon, whereas another coincidence $r_b=r_i$ is understood as the extremal configuration for black holes in KNTNdS spacetime. Some illustrations of these extremal states are given in Figs. \ref{fig.extreme}, \ref{fig.extremeQ}, and \ref{fig.extremen}. These plots confirm several extremalities that can exist in KNTNdS spacetime. 

The area of cosmological horizon can be computed by using the standard formula
\be 
A_c  = \int\limits_{\phi  = 0}^{2\pi } {\int\limits_{\theta  = 0}^\pi  {\sqrt {g_{\theta \theta } g_{\phi \phi } } d\theta d\phi } }  = 4\pi \left( {r_c^2  + a^2 + n^2 } \right)\,,
\ee 
whereas the surface gravity associated to the cosmological horizon can be given by \cite{Wu:2019pzr,Wu:2022rmx}
\be 
\kappa_c  = \frac{{2\pi }}{{A_c }}\left. {\frac{{d\Delta _r }}{{dr}}} \right|_{r_c } \,,
\ee 
which gives
\be 
\kappa_c  = \frac{{\left( {r_c  - M} \right) l^2  - r_c \left( {2r_c^2  + 6n^2  + a^2 } \right)}}{{\left( {r_c^2  + a^2 + n^2 } \right)l^2  }}\,.
\ee 
Accordingly, the Hawking temperature that corresponds to the cosmological horizon can be computed by using the relation $T_c = \kappa_c/2\pi$. The angular velocity and electric potential at the cosmological horizon can be written as
\be 
\Omega _c  = \frac{a}{{r_c^2  + a^2  + n^2 }}\,,
\ee 
and
\be 
\Phi _c  = \frac{{qr_c }}{{r_c^2  + a^2  + n^2 }}\,,
\ee 
respectively. 

The thermodynamics of Taub-NUT spacetime has become a lively discussion in recent years \cite{Wu:2019pzr,Wu:2022rmx,Awad:2022jgn,BallonBordo:2020mcs,Rodriguez:2021hks,Pradhan:2020ofm}. One of the proposals is to associate an entropy for the Misner string in the Taub-NUT spacetime \cite{BallonBordo:2020mcs}. Another is to introduce some conserved charges, i.e. $\Xi = Mn$ and $N=n$, so the resulting first law of horizon mechanics would read \cite{Wu:2019pzr}
\be \label{eq.firstlaw}
\delta M = T\delta S + \Omega \delta J + \Phi \delta Q + \Omega_n \delta J_n  + \Psi \delta N + \Theta \delta \Lambda \,.
\ee 
The last term in eq. (\ref{eq.firstlaw}) represents the change of total energy inside of the cosmological horizon due to the variation of cosmological constant. In section \ref{s4.microentropy} where we compute the microscopic formula for the entropy, eq. (\ref{eq.firstlaw}) will be used to obtain the corresponding conjugate charge. Note that in the vanishing of $\delta \Lambda$, $\delta N$, and $\delta J_n$, this first law of horizon mechanics simply reduces to the one which is related to the Kerr-Newman black hole \cite{Chen:2010ywa}.

\section{Hidden conformal symmetries}\label{s3.hidden}

The hidden conformal symmetry can be shown from the corresponding massless test scalar field in the near region of cosmological horizon. Here we consider the test charged massless scalar with an equation of motion
\be \label{KGeqtn}
\left( {\nabla _\mu   + iqA_\mu  } \right)\left( {\nabla ^\mu   + iqA^\mu  } \right)\Psi  = 0\,.
\ee
Note that the KNTNdS spacetime possesses the stationary and axial Killing symmetry, which allow us to express the test scalar wave function to be separable in the following way,
\be
\Psi  = e^{ - i\omega t + im\phi } X\left( x \right)R\left( r \right)\,.
\ee
Using this ansatz, the Klein-Gordon equation above can be written as
\[
\frac{1}{X\left( x \right)}\frac{d}{{dr}}\left( {\Delta _r \frac{{dR\left( r \right)}}{{dr}}} \right) + \frac{1}{R\left( r \right)}\frac{d}{{dx}}\left( {\Delta _x \Delta _l \frac{{dX\left( x \right)}}{{dx}}} \right)
\]
\be
+ \left\{ {\frac{{\left( {\left( {r^2  + a^2  + n^2 } \right)\omega  + qQr - ma} \right)^2 }}{{\Delta _r }} }  - \frac{{\left( {ax^2  + 2nx\omega  - am\omega } \right)^2 }}{{\Delta _l \Delta _x }} \right\}  = 0\,.
\ee
It turns out that the last equation can be separated into two equations, namely the radial part
\be \label{eq.rad}
\frac{d}{{dr}}\left( {\Delta _r \frac{{dR\left( r \right)}}{{dr}}} \right) + \left( {\frac{{\left( {\left( {r^2  + a^2  + n^2 } \right)\omega  + qQr - ma} \right)^2 }}{{\Delta _r }} - \lambda } \right)R\left( r \right) = 0\,,
\ee
and the angular one
\be \label{eq.ang}
\frac{d}{{dx}}\left( {\Delta _x \Delta _l \frac{{dX\left( x \right)}}{{dx}}} \right) - \left(   - \frac{{\left( {ax^2  + 2nx\omega  - am\omega } \right)^2 }}{{\Delta _l \Delta _x }} - \lambda  \right)X\left( x \right) = 0\,,
\ee
with $\lambda$ as the separable constant.

The hidden conformal symmetry is related to the radial equation (\ref{eq.rad}) under some circumstances. However, before we proceed to show the hidden conformal symmetry, we need to consider an approximation to the $\Delta_r$ in the radial equation (\ref{eq.rad}) so it can be quadratic in $r$. Such approximation has been repeatedly performed for the hidden conformal symmetry in the (anti)-de Sitter spacetime \cite{Sakti:2019zix,Sakti:2020jpo,Chen:2010bh}. The approximation is basically a Taylor expansion for $\Delta_r$ near the cosmological horizon radius,
\be
\Delta _r  \simeq \left. {\Delta _r } \right|_{r_c }  + \left. {\frac{{d\Delta _r }}{{dr}}} \right|_{r_c } \left( {r - r_c } \right) + \left. {\frac{{d^2 \Delta _r }}{{dr^2 }}} \right|_{r_c } \frac{{\left( {r - r_c } \right)^2 }}{2}\,,
\ee
that can give us
\be
\Delta _r  \simeq K\left( {r - r_c } \right)\left( {r - r_c^* } \right)\,.
\ee
Related to the KNTNdS spacetime (\ref{metricKNTNdS}), we have
\be
K = 1 - l^{ - 2} \left( {a^2  + 6n^2  + 6r_c^2 } \right)\,,
\ee
and
\be
r_c^*  = \frac{{2r_c^3  + r_c \left( {l^2  - a^2  - 6n^2 } \right) - 2Ml^2 }}{{a^2  + 6n^2  + 6r_c^2  - l^2 }}\,.
\ee
Now let us impose the near horizon region for the test scalar, given by $\omega r \ll 1$. We also consider that the test scalar is in the low frequency, $\omega M \ll 1$, and its charge is extremely small, $qQ \ll 1$. Subsequently, the low frequency can be related to the conditions $\omega a \ll 1$, $\omega Q \ll 1$, and $\omega n \ll 1$.

Under such considerations, the radial equation (\ref{eq.rad}) can be rewritten approximately as
\be \label{eq.rad.app}
\frac{d}{{dr}}\left( {\left( {r - r_c } \right)\left( {r - r_c^* } \right)\frac{{dR}}{{dr}}} \right) + {\frac{{F_1 R}}{{\left( {r - r_c } \right)}} + \frac{{F_2 R}}{{\left( {r - r_c^* } \right)}}} + F_3 R = 0\,,
\ee
with
\be
F_1  = \frac{{\left( {\omega \left( {r_c^{2}  + a^2 + n^2 } \right)+ qQr_c - ma  } \right)^2 }}{{K^2 \left( {r_c  - r_c^* } \right) }}\,,
\ee
\be
F_2  =  - \frac{{\left( {\omega \left( {r_c^{*2}  + a^2 + n^2 } \right) + qQr_c^*- ma   } \right)^2 }}{{K^2 \left( {r_c  - r_c^* } \right) }}\,,
\ee
and
\be
F_3  = \frac{{q^2 Q^2 -\lambda }}{{K }}\,.
\ee

Furthermore, to show the symmetry property of eq. (\ref{eq.rad.app}), we consider the following conformal coordinates
\be\label{w0}
\omega^ +   = \sqrt {\frac{{r - r_c }}{{r - r_c^* }}} \exp \left( {2\pi T_R \phi  + 2n_R t} \right)\,,
\ee
\be \label{wp}
\omega^ +   = \sqrt {\frac{{r - r_c }}{{r - r_c^* }}} \exp \left( {2\pi T_L \phi  + 2n_L t} \right)\,,
\ee
\be \label{wm}
\omega^0  = \sqrt {\frac{{r_c  - r_c^* }}{{r - r_c^* }}} \exp \left( {\pi \left( {T_R  + T_L } \right)\phi  + \left( {n_R  + n_L } \right)t} \right)\,.
\ee
By using these coordinates, and also the notations
\be
\partial _{+ }  = \frac{\partial }{{\partial \omega^ +  }}~~,~~\partial _{ - }  = \frac{\partial }{{\partial \omega^ -  }}~~,~~\partial _{0}  = \frac{\partial }{{\partial \omega^0 }}\,,
\ee
one can define two copies of $SL(2,\mathbb{R})$ generators that read
\be
H_{+ }  = i\partial _{+ } \,,
\ee
\be
H_{0}  = i\left( {\omega^ +  \partial _{k, + }  + \frac{1}{2}\omega^0 \partial _{0} } \right)\,,
\ee
\be
H_{- }  = i\left( {\left( {\omega^ +  } \right)^2 \partial _{ + }  + \omega^ +  \omega^0 \partial _{0}  - \left( {\omega^0 } \right)^2 \partial _{- } } \right)\,,
\ee
and
\be
\bar H_{+ }  = i\partial _{ - } \,,
\ee
\be
\bar H_{0}  = i\left( {\omega^ -  \partial _{- }  + \frac{1}{2}\omega^0 \partial _{0} } \right)\,,
\ee
\be
\bar H_{- }  = i\left( {\left( {\omega^ -  } \right)^2 \partial _{- }  + \omega^ -  \omega^0 \partial _{0}  - \left( {\omega^0 } \right)^2 \partial _{ + } } \right)\,.
\ee
The generators above satisfy the  $SL(2,\mathbb{R})$ algebra,
\be
\left[ {H_{0} ,H_{ \pm } } \right] =  \mp iH_{ \pm }~~,~~ \left[ {H_{ + } ,H_{ - } } \right] = 2iH_{0} \,,
\ee
and
\be
\left[ {\bar H_{0} ,\bar H_{ \pm } } \right] =  \mp i\bar H_{ \pm } ~~,~~\left[ {\bar H_{ + } ,\bar H_{ - } } \right] = 2i\bar H_{0} \,.
\ee

Moreover, based on the generators above, one can also construct the $SL(2,\mathbb{R})$ Casimir operators
\be
{\cal H}^2  =  - H_{0}^2  + \frac{1}{2}\left( {H_{ + } H_{ - }  + H_{ - } H_{ + } } \right)\,,
\ee
and
\be
\bar {\cal H}^2  =  - \bar H_{0}^2  + \frac{1}{2}\left( {\bar H_{ + } \bar H_{ - }  + \bar H_{ - } \bar H_{ + } } \right)\,.
\ee
These Casimir operators commute with all generators in the group. It turns out that despite the Casimir operators ${\cal H}^2$ and $\bar{\cal H}$ are constructed by a different set of operators, they take exactly the same form in terms of the conformal coordinates
\be \label{Casimir}
{\cal H}^2  = \bar {\cal H}^2  = \frac{1}{4}\left( {\left( {\omega ^0 } \right)^2 \partial _{0}^2  - \omega ^0 \partial _{0} } \right) + \left( {\omega ^0 } \right)^2 \partial _{ + } \partial _{ - }
\ee
Furthermore, in terms of the coordinates $(t,r,\phi)$, the Casimir operator (\ref{Casimir}) explicitly can be written as
\[
{\cal H}^2  = \left( {r - r_c } \right)\left( {r - r_c^* } \right)\frac{{\partial ^2 }}{{\partial r^2 }} + \left( {2r - r_c^*  - r_c } \right)\frac{\partial }{{\partial r}} + \frac{{\left( {r_c  - r_c^* } \right)}}{{r - r_c^* }}\left( {\frac{{n_L  - n_R }}{{4\pi {\Xi} }}\frac{\partial }{{\partial \phi }} - \frac{{T_L  - T_R }}{{4{\Xi} }}\frac{\partial }{{\partial t}}} \right)^2
\]
\be \label{eq.Casimir}
- \frac{{\left( {r_c  - r_c^* } \right)}}{{r - r_c }}\left( {\frac{{n_L  + n_R }}{{4\pi {\Xi} }}\frac{\partial }{{\partial \phi }} - \frac{{T_L  + T_R }}{{4{\Xi} }}\frac{\partial }{{\partial t}}} \right)^2 \,,
\ee
where $\Xi = n_L T_R - n_R T_L$. The hidden conformal symmetry is stated from the fact that the radial equation (\ref{eq.rad.app}) is just the eigen equation for the Casimir operator ${\cal H}^2$ in eq. (\ref{eq.Casimir}) with the separation constant $\lambda$ as the corresponding eigenvalue. In this fashion, it is said that the radial equation (\ref{eq.rad.app}) exhibits the $SL(2,\mathbb{R})_L \times SL(2,\mathbb{R})_R$ symmetry.

In the development, the discussion of hidden conformal symmetry for the black hole/CFT correspondence was found to exist in several pictures \cite{Chen:2010ywa,Chen:2011kt}. The first one is known as the $J$-picture which uses the neutral test scalar to probe the symmetry. This is actually the original proposal of hidden conformal symmetry in Kerr/CFT correspondence \cite{Castro:2010fd}, where the property of low frequency neutral test scalar in the near region of Kerr black hole is explored. The second one is called $Q$-picture \cite{Chen:2010as}, where a $U(1)$ internal space is introduced and the test scalar is considered to be in the state of zeroth quantum magnetic number. Subsequently, these $J$ and $Q$ pictures can be formulated into a single treatment known as the general picture \cite{Chen:2011kt} where the CFT duals are generated by the modular group $SL(2,{\mathbb Z})$. In the following we show that the twofold or even the general pictures of hidden conformal symmetry that are applied for black hole horizon \cite{Chen:2010ywa,Chen:2011kt} can also be found near the cosmological horizon.

The $J$ picture is obtained by considering a neutral scalar test probe in the near horizon region. This consideration yields eq. (\ref{eq.rad.app}) reduces to
\be \label{eq.rad.app.Jpic}
\frac{d}{{dr}}\left( {\left( {r - r_c } \right)\left( {r - r_c^* } \right)\frac{{dR}}{{dr}}} \right) + {\frac{{G_1^J R}}{{\left( {r - r_c } \right)}} + \frac{{G_2^J R}}{{\left( {r - r_c^* } \right)}}} + G_3^J R = 0\,,
\ee
where
\be \label{C1J}
G_1^J  = \frac{{\left( {\omega \left( {r_c^{*2}  + a^2 + n^2 } \right) - ma } \right)^2 }}{{K^2 \left( {r_c  - r_c^* } \right) }}\,,
\ee
\be \label{C2J}
G_2^J  =  - \frac{{\left( {\omega \left( {r_c^{*2}  + a^2 + n^2} \right) - ma   } \right)^2 }}{{K^2 \left( {r_c  - r_c^* } \right) }}\,,
\ee
and
\be \label{C3J}
G_3^J  = - \frac{\lambda }{K}\,.
\ee
To have an agreement between the radial equation (\ref{eq.rad.app.Jpic}) with the general form (\ref{eq.Casimir}), we need to set the corresponding parameters as
\be
n_R^J  = 0~~,~~n_L^J  = - \frac{K}{{2\left( {r_c  + r_c^* } \right)}}\,,
\ee
and
\be \label{TLTRJ}
T_R^J  = \frac{{K \left( {r_c  - r_c^* } \right)}}{{4\pi a}}~~,~~T_L^J  = \frac{{K \left( {r_c^2  + {r_c^*}^2  + 2a^2 + 2n^2 } \right)}}{{4\pi a \left( {r_c  + r_c^* } \right)}}\,.
\ee
In this way, we have constructed the $J$ picture hidden conformal symmetry near the cosmological horizon of the KNTNdS spacetime.

If the $J$-picture corresponds to the neutral scalar test field, the $Q$-picture can be probed by using the scalar field at $m=0$ state. However, the $Q$-picture hidden conformal symmetry requires the existence of a $U(1)$ internal space $\chi$. The dependence of scalar wave $\Psi$ with respect to this new coordinate is given by the eigen equation
\be
\partial _\chi  \Psi  = i q \Psi \,.
\ee
Indeed, it has an analogy with the equation for the angular coordinate $\phi$ where the scalar probe dependence is given by $\partial _\phi  \Psi  = i m \Psi$. For the $m=0$ state, the radial equation (\ref{eq.rad.app}) takes the form
\be \label{eq.rad.app.Qpic}
\frac{d}{{dr}}\left( {\left( {r - r_c } \right)\left( {r - r_c^* } \right)\frac{{dR}}{{dr}}} \right) + {\frac{{G_1^Q R}}{{\left( {r - r_c } \right)}} + \frac{{G_2^Q R}}{{\left( {r - r_c^* } \right)}}}+ G_3^Q R = 0\,,
\ee
where
\be \label{C1Q}
G_1^Q  = \frac{{\left( {\omega \left( {r_c^{2}  + a^2 +n^2} \right) + qQr_c } \right)^2 }}{{K^2 \left( {r_c  - r_c^* } \right) }}\,,
\ee
\be \label{C2Q}
G_2^Q  =  - \frac{{\left( {\omega \left( {r_c^{*2}  + a^2+n^2 } \right)  + qQr_c^* } \right)^2 }}{{K^2 \left( {r_c  - r_c^* } \right) }}\,,
\ee
and
\be \label{C3Q}
G_3^Q  = \frac{{q^2 Q^2 }}{{K^2 }}\,.
\ee

In order to have an agreement between (\ref{eq.Casimir}) and (\ref{eq.rad.app.Qpic}), the corresponding parameters should be set as
\be
n_L^Q  =  - \frac{{K\left( {r_c  + r_c^* } \right)}}{{4 \left( {r_cr_c^*  - a^2-n^2 } \right) }}~~,~~n_R^Q  =- \frac{{K\left( {r_c  - r_c^* } \right)}}{{4 \left( {r_cr_c^*  - a^2-n^2 } \right) }}\,,
\ee
and
\be \label{TLTRQ}
T_L^Q  = -\frac{{K \left( {r_c^2  + r_c^{*2} +2 a^2 + 2 n^2 } \right)}}{{4\pi Q\left( {r_c r_c^*  - a^2 -n^2} \right)}}~~,~~T_R^Q  = -\frac{{K \left( {r_c^2  - r_c^{*2} } \right)}}{{4\pi Q\left( {r_c r_c^*  - a^2 - n^2 } \right)}}\,.
\ee
Here we have shown that the $Q$-picture of hidden conformal symmetry for the cosmological horizon of KNTNdS spacetime does exist as well.

Now let us discuss the general picture which combines both $J$ and $Q$-pictures in a single formulation. The similarity between eqs. (\ref{eq.rad.app.Jpic}) and (\ref{eq.rad.app.Qpic}) are obvious. It then allows us to write an equation that utilizes the $SL(2,{\mathbb Z})$ modular group which can reduce to both eqs. (\ref{eq.rad.app.Jpic}) and (\ref{eq.rad.app.Qpic}) for some special cases. The $SL(2,{\mathbb Z})$ modular group itself acts on the torus with coordinates $(\phi, \chi)$ that appear in eqs. (\ref{eq.rad.app.Jpic}) and (\ref{eq.rad.app.Qpic}). The corresponding $SL(2,{\mathbb Z})$ transformation is given by
\be
\left( {\begin{array}{*{20}c}
		{\phi '}  \\
		{\chi '}  \\
\end{array}} \right) = \left( {\begin{array}{*{20}c}
		\alpha  & \beta   \\
		\gamma  & \delta   \\
\end{array}} \right)\left( {\begin{array}{*{20}c}
		\phi   \\
		\chi   \\
\end{array}} \right)\,.
\ee
Intuitively, related to the new coordinates $\phi '$ and $\chi '$, the corresponding eigenvalues of operators $i\partial_{\phi '}$ and $i\partial_{\chi '}$ are $m'$ and $q'$, respectively. Therefore, one can write the relation between the parameters $(m,q)$ and $(m',q')$ as
\be
m = \alpha m' + \gamma q'~~,~~q = \beta m' + \delta q'\,.
\ee

In terms of $(m',q')$ parameters, the radial equation (\ref{eq.rad.app}) can be rewritten as

\be \label{eq.rad.app.gen}
\frac{d}{{dr}}\left( {\left( {r - r_c } \right)\left( {r - r_c^* } \right)\frac{{dR}}{{dr}}} \right) + {\frac{{G_1^G R}}{{\left( {r - r_c } \right)}} + \frac{{G_2^G R}}{{\left( {r - r_c^* } \right)}}} + G_3^G R = 0\,,
\ee
where
\be
G_1^G  = \frac{{\left( {\omega \left( {r_c^{2}  + a^2 + n^2 } \right) - \left( {a\alpha  - Q\beta r_c } \right)m' - \left( {a\gamma  - Q\delta r_c } \right)q'  } \right)^2 }}{{K^2 \left( {r_c  - r_c^* } \right) }}\,,
\ee
\be
G_2^G  =  - \frac{{\left( {\omega \left( {r_c^{*2}  + a^2 + n^2 } \right)  - \left( {a\alpha  - Q\beta r_c^* } \right)m' - \left( {a\gamma  - Q\delta r_c^* } \right)q'   } \right)^2 }}{{K^2 \left( {r_c  - r_c^* } \right) }}\,,
\ee
and
\be
G_3^G  = \frac{{(\beta m' + \delta q')^2 Q^2 -\lambda }}{{K }}\,.
\ee
In this general picture, the $J'$-picture can be obtained by setting $q' = 0$, and the $Q'$-picture is achieved at $m'=0$. Note that these two $J'$ and $Q'$ pictures are equal to each other. In $Q'$-picture, the hidden conformal symmetry exists if we set the corresponding parameters in eq. (\ref{eq.Casimir}) to be
\be
n_L^G  =  - \frac{{K\left( {2a\gamma  - \delta Q\left( {r_c  + r_c^* } \right)} \right)}}{{4\left( {a\gamma \left( {r_c  + r_c^* } \right) + \delta Q\left( {a^2  + n^2  - r_c r_c^* } \right)} \right)}}\,,
\ee
\be
n_R^G  = \frac{{K\delta Q\left( {r_c  - r_c^* } \right)}}{{4\left( {a\gamma \left( {r_c  + r_c^* } \right) + \delta Q\left( {a^2  + n^2  - r_c r_c^* } \right)} \right)}}\,,
\ee
\be
T_L^G  = \frac{{K\left( {r_c^2  + \left( {r_c^* } \right)^2  + 2a^2  + 2n^2 } \right)}}{{4\pi \left( {a\gamma \left( {r_c  + r_c^* } \right) + \delta Q\left( {a^2  + n^2  - r_c r_c^* } \right)} \right)}}\,,
\ee
and
\be
T_R^G  = \frac{{K\left( {r_c^2  - \left( {r_c^* } \right)^2 } \right)}}{{4\pi \left( {a\gamma \left( {r_c  + r_c^* } \right) + \delta Q\left( {a^2  + n^2  - r_c r_c^* } \right)} \right)}}\,.
\ee

\section{Microscopic entropy}\label{s4.microentropy}

In the previous section, we have established the hidden conformal symmetries that associate to the cosmological horizon in KNTNdS spacetime. These symmetries open the possibility to apply some CFT$_2$ methods to understand several aspects related to the cosmological horizon. In this section, we show that the cosmological horizon entropy has a dual description from a CFT$_2$ point of view by using the Cardy formula. The same argument has been used to construct Kerr/CFT holography to recover the entropy of black holes \cite{Guica:2008mu,Compere:2012jk}. The dual calculations for cosmological horizon entropy are performed separately in $J$ and $Q$-pictures where each of them requires a particular approach to obtain the corresponding near horizon geometry.

Let us first consider the $J$-picture. The appropriate near horizon coordinate transformation in the near extremal state is
\be \label{eq.transNearKerrCFT}
r = \frac{{r_c  + r_c^* }}{2} + \varepsilon r_0 \tilde r~~,~~r_c  - r_c^*  = \mu \varepsilon r_0 ~~,~~t = \tilde t\frac{{r_0}}{\varepsilon }~~,~~\phi  = \tilde \phi  + \Omega _H \tilde t \frac{{r_0}}{\varepsilon }\,.
\ee 
Applying this transformation to the metric (\ref{metricKNTNdS}) gives
\be \label{metric.nearJ}
ds^2  = \Gamma \left( \theta  \right)\left\{ { - {\left( {\tilde r - \frac{{\mu }}{2}} \right)\left( {\tilde r + \frac{{\mu }}{2}} \right) } d\tilde t^2  + \frac{{d\tilde r^2 }}{\left( {\tilde r - \frac{{\mu }}{2}} \right)\left( {\tilde r + \frac{{\mu }}{2}} \right)} + \alpha \left( x  \right)dx^2 } \right\} + \gamma \left( \theta  \right)\left( {d\tilde \phi  + \tilde k^J {\tilde r} d\tilde t} \right)^2 \,,
\ee 
where
\be
\Gamma \left( \theta  \right) = \frac{{\rho _ c ^2 }}{{K}}\,,
\ee
\be
\alpha \left( x \right) = \frac{{K }}{{\Delta _l \Delta_x }}\,,
\ee
\be
\gamma \left( \theta  \right) = \frac{{\Delta _l \Delta_x \left( {r_c^2  + a^2 + n^2 } \right)^2 }}{{\rho _c^2 }}\,,
\ee
\be
\tilde k^J = \frac{{2 ar_c }}{{ \left( {r_ c ^2  + a^2 + n^2 } \right) K }}\,,
\ee
and
\be
\rho _c^2  = r_c^2  + a^2 x^2 \,.
\ee

A general formula to compute the central charge associated to the near horizon of a near extremal geometry in the family of Einstein-Maxwell theory has been computed in \cite{Hartman:2008pb}, and the obtained formula can apply to the near horizon geometry that appears in eq. (\ref{metric.nearJ}). The central charges are \cite{Hartman:2008pb}
\be \label{c.formula}
c_L^J  = c_R^J  = 3\tilde k^J \int\limits_{-1}^1  {dx \sqrt {\Gamma \left( x \right)\gamma \left( x  \right)\alpha \left( x  \right)} } \,,
\ee  
for the cosmological horizon in KNTNdS spacetime we have
\be \label{centralchargeJ}
c_L^J  = c_R^J = \frac{6a(r_c + r_c^*)}{K}\,.
\ee 
Using the central charge (\ref{centralchargeJ}), the Cardy formula
\be 
S_{Cardy}^J = \frac{{\pi ^2 }}{3}\left( {c_L^J T_L^J  + c_R^J T_R^J } \right)
\ee 
gives us the entropy of cosmological horizon in KNTNdS spacetime, namely
\be \label{entropy}
S_{Cardy}^J = {\pi \left( {r_c^2  + a^2 +n^2 } \right)}\,.
\ee 

In the next section, we will provide the holographic calculation for absorption cross section by using a two point function of the operator that is dual to the scalar field.  For that calculation, one needs to identify the conjugate charges $\delta E_L^J$ and $\delta E_R^J$ that satisfy
\be 
\delta S_{Cardy}^J  = \frac{{\delta E_L^J }}{{T_L^J }} + \frac{{\delta E_R^J }}{{T_R^J }}\,.
\ee 
The conjugate charges can be obtained by using the first law of thermodynamics for the cosmological horizon (\ref{eq.firstlaw}) and the corresponding equation is
\be\label{eq.charges}
\delta S  = \frac{{\delta M  - \Omega \delta J - \Phi \delta Q - \Psi dN - \Omega_n \delta J_n  - \Theta  \delta \Lambda  }}{{T}} = \frac{{\delta E_L^J }}{{T_L^J }} + \frac{{\delta E_R^J }}{{T_R^J }}\,.
\ee 
Recall that for the Kerr-Newman black hole, the related first law of thermodynamics can be written as
\be \label{eq.deltaMnormal}
\delta M = T\delta S + \Omega \delta J + \Phi \delta Q\,.
\ee 
To obtain the conjugate charges in Kerr/CFT correspondence for the Kerr-Newman black hole, the variations of black hole parameters in eq. (\ref{eq.deltaMnormal}) are related to some of the scalar probe parameters. The change in angular momentum is related to the magnetic quantum number $m$, whereas the change in electric charge is connected with the probe's electric charge. 

However, the first law of thermodynamics (\ref{eq.charges}) contains several terms that do not have a direct counterpart with the scalar probe's properties. Therefore, here we propose some generalizations for the variations related to the rotational and electromagnetic works, i.e.
\be \label{Omprimed}
\Omega ' \delta J'  = \Omega \delta J + \Omega _n \delta J_n  + \Theta \delta \Lambda \,.
\ee 
and
\be \label{Qprimed}
\Phi ' \delta Q'  = \Phi  \delta Q + \Psi \delta N \,,
\ee 
respectively. In such generalizations, we can have $\delta M = \omega$ and $\delta J' = m$. In $J$ picture, we can have
\be \label{dEJ}
\delta E_L^J  = \omega _L^J  ~~{\rm and}~~\delta E_R^J  = \omega _R^J  \,,
\ee 
where
\be \label{wLRJ}
\omega _L^J  = \frac{{r_c^2  + \left( {r_c^* } \right)^2  + 2a^2 + 2n^2 }}{{2a }}\omega
~~,~~
\omega _R^J  = \frac{{r_c^2  + \left( {r_c^* } \right)^2  + 2a^2 + 2n^2 }}{{2a }}\omega - m\,,
\ee
\be \label{qLJ}
q_L^J = q_R^J=0\,,~~{\rm and}~~\mu_L^J = \mu_R^J = 0\,.
\ee

Now let us turn to microscopic entropy calculation associated with the $Q$-picture hidden conformal symmetry. As it has been mentioned previously, we need to add a $U(1)$ internal extra dimension in the theory denoted by the coordinate $\chi$. The corresponding five dimensional spacetime metric then can be read as
\be 
ds^2  = ds_4^2  + \left( {d\chi  + \tilde {\bf A}} \right)^2 \,,
\ee 
where $ds_4^2$ is the near horizon metric of near extremal horizon (\ref{metric.nearJ}) and $\tilde {\bf A}$ is the associated gauge field. Now let us consider the fluctuation of five dimensional metric with a set of coordinate $\{ \tilde t,\tilde \phi,\theta ,\tilde r ,\chi \} $ in the following way,
\be \label{eq.BCqpic}
h_{\mu \nu }  \sim \left( {\begin{array}{*{20}c}
		{{\cal O}\left( {\tilde r^2 } \right)} & {{\cal O}\left( {\tilde r} \right)} & {{\cal O}\left( {1/\tilde r} \right)} & {{\cal O}\left( {1/\tilde r^2 } \right)} & {{\cal O}\left( 1 \right)}  \\
		{} & {{\cal O}\left( {1/\tilde r} \right)} & {{\cal O}\left( 1 \right)} & {{\cal O}\left( {1/\tilde r} \right)} & {{\cal O}\left( 1 \right)}  \\
		{} & {} & {{\cal O}\left( {1/\tilde r} \right)} & {{\cal O}\left( {1/\tilde r^2 } \right)} & {{\cal O}\left( {1/\tilde r} \right)}  \\
		{} & {} & {} & {{\cal O}\left( {1/\tilde r^3 } \right)} & {{\cal O}\left( {1/\tilde r} \right)}  \\
		{} & {} & {} & {} & {{\cal O}\left( 1 \right)}  \\
\end{array}} \right)\,.
\ee  
The most general diffeomorphism that preserves the boundary condition in (\ref{eq.BCqpic}) is
\be 
\zeta ^{\left( \chi  \right)}  = \varepsilon \left( \chi  \right)\frac{\partial }{{\partial \chi }} - \tilde r\frac{{d\varepsilon \left( \chi  \right)}}{{d\chi }}\frac{\partial }{{\partial \tilde r}}\,,
\ee 
where $\varepsilon \left( \chi  \right) = \exp \left(-iq\chi\right)$. To compute the central charge, we can generalize the treatment developed in \cite{Hartman:2008pb} and we can get
\be \label{centralchargeQ}
c_L^Q  = c_R^Q  = 3 {\tilde k^Q} \int\limits_{-1}^{1}  {dx \sqrt {\Gamma \left( x  \right)\gamma \left( x  \right)\alpha \left( x  \right)} } = \frac{6Q \left(r_c r_c^* -a^2 - n^2\right)}{K}\,.
\ee  
Note that the negative value of this central charge leads to the non-unitary CFT$_2$. Therefore, to guarantee the unitary, we have to impose $r_c r_c^* > a^2 +n^2$ and the positive values consideration of $Q$ and $K$. However, sometime the non-unitary dual CFT$_2$ is unavoidable in several systems. For example, in the case of a strong magnetic field for the magnetized Kerr/CFT holography \cite{Siahaan:2015xia}.
This central charge then can be used to compute the microscopic entropy by using Cardy formula
\be 
S = \frac{{\pi ^2 }}{3}\left( {c_L^Q T_L^Q  + c_R^Q T_R^Q } \right)\,,
\ee 
which recovers the entropy of the cosmological horizon in KNTNdS spacetime in eq. (\ref{entropy}). 

Similar to the $J$-picture, the conjugate charges in $Q$-picture can be extracted from the first law of thermodynamics, i.e.
\be\label{eq.chargesQ}
\delta S  = \frac{{\delta M  - \Omega' \delta J' - \Phi ' \delta Q'  }}{{T }} = \frac{{\delta E_L^Q }}{{T_L^Q }} + \frac{{\delta E_R^Q }}{{T_R^Q }}\,,
\ee 
where $\Omega' \delta J'$ is given in (\ref{Omprimed}) and $\Phi ' \delta Q'$ is given in (\ref{Qprimed}). However, since in the $Q$ picture we consider the case of $m=0$ only, the term $\Omega' \delta J'$ can be neglected. The last equation can be solved by setting
\be 
\delta E_L^Q  = \omega _L^Q  - q_L^Q \mu _L^Q    ~~,~~\delta E_R^Q  = \omega _R^Q  - q_R^Q \mu _R^Q \,,
\ee
where
\be\label{wLRQ}
\omega _L^Q  = \omega _R^Q  = \frac{{n\left( {r_c  + r_c^* } \right)\left( {r_c^2  + \left( {r_c^* } \right)^2  + 2a^2 + 2n^2} \right)}}{{2Q\left( {r_c r_c^*  - a^2 - n^2} \right)}}\omega  \,,
\ee
\be\label{qLRQ}
q_L^Q  = q_R^Q  = \delta Q_c  = q\,,
\ee
and
\be\label{muLRQ}
\mu _L^Q  = \frac{{n\left( {r_c^2  + \left( {r_c^* } \right)^2  + 2a^2+2n^2 } \right)}}{{2\left( {r_c r_c^*  - a^2 -n^2 } \right)}}~~,~~\mu _R^Q  = \frac{{n\left( {r_c  + r_c^* } \right)^2 }}{{2\left( {r_c r_c^*  - a^2-n^2 } \right)}}\,.
\ee

Before proceeding to the next section, let us here give some remarks on the results above. As we have mentioned in the introduction that a holography for de Sitter geometry has been reported in \cite{Strominger:2001pn} where the author established a connection between the gravity theory in D-dimensional de Sitter space with a conformal field theory living in (D-1)-sphere. Particularly, for the three dimensional de Sitter (dS$_3$) background, the corresponding dual theory is a two dimensional conformal field theory which is exactly similar to the case presented in this paper. Moreover, the technique to compute the central charge that associates to the asymptotics symmetries of dS$_3$ is the analogous calculation by Brown and Henneaux in \cite{Brown:1986nw}. In fact, the method by Brown and Henneaux for the central charge is also the method adopted in Kerr/CFT correspondence \cite{Guica:2008mu}. As we have mentioned previously, the work presented in this paper is an application of Kerr/CFT correspondence to the cosmological horizon of KNTNdS spacetime. Therefore, one can say that basically the work in this paper and the dS/CFT correspondence reported in \cite{Strominger:2001pn} depart from the same basis. Note that the author of \cite{Strominger:2001pn} is also one of the authors of \cite{Guica:2008mu} and \cite{Castro:2010fd} that establish the Kerr/CFT correspondence.

However, the constructions of dS/CFT holography in \cite{Strominger:2001pn} and the one presented here are distinguished in several aspects. First, the dual CFT theory proposed in \cite{Strominger:2001pn} is (D-1)-dimensional, for D as the dimension of gravitational theory in dS$_3$. Typically in the Kerr/CFT holography, the dual CFT theory is two dimensional regardless of the dimension of bulk spacetime under consideration \cite{Strominger:2001pn,Compere:2012jk}. Second, the asymptotic geometry considered in \cite{Strominger:2001pn} is still de Sitter, whereas in this paper the corresponding near-horizon of (near)-extremal geometry as appeared in eq. (\ref{metric.nearJ}) is warped anti-de Sitter. Consequently, these distinguished asymptotic geometries require different types of boundary conditions to be used in the diffeomorphism generators that lead to distinctive central charges. Third, the central charge obtained in this work fails to reduce smoothly to the non-rotating or neutral cases. It can be observed that in general the central charges (\ref{centralchargeJ}) and (\ref{centralchargeQ}) vanish in the case of $a=0$ and $Q=0$, respectively. This limitation is well known for the Kerr/CFT correspondence as it works for the rotating and/or charged case only, i.e. the black hole/CFT correspondence for Schwarzschild black hole in the style of \cite{Guica:2008mu} is not well established. On the other hand, the holography for de Sitter spacetime established in \cite{Strominger:2001pn} can be extended to the case that incorporates black holes \cite{Klemm:2001ea}.

\section{CFT$_2$ correlators and superradiant scattering}\label{s5.scattering}

Normally, in the Kerr/CFT correspondence discussions for black holes, another evidence in supporting the holography is the dual calculation for scalar absorption near the horizon. Similar to that, here we also provide such calculations in the near cosmological horizon, and show that the dictionary of Kerr/CFT correspondence still applies. To start, let us consider the frequency near superradiant bound for the scalar field
\be
\omega - m\Omega - q \Phi = {\bar \omega} \frac{\varepsilon}{r_0}\,.
\ee
The scattering analysis can be done in $J$ or $Q$ picture, and consequently it should be feasible in the general picture as well. Here let us consider the $J$-picture only, where the corresponding radial equation is eq. (\ref{eq.rad.app.Jpic}). The solution to this equation reads
\be \label{eq.RadSolScattering}
R\left( \zeta \right) = \zeta^\alpha  \left( {1 - \zeta} \right)^\beta  F\left( {a,b,c;\zeta} \right)\,,
\ee
where $\zeta = \left( {r - r_c } \right)\left( {r - r_c^* } \right)^{ - 1} $,
\be
i\alpha  = \sqrt {G_1^J } \,,
\ee
\be
2 \beta  = 1 - \sqrt {1 - 4 G_3^J } \,,
\ee
and
\be
a^{*}  = \alpha  + \beta  + i\sqrt { - G_2^J }~~,~~ b^{*}  = \alpha  + \beta  - i\sqrt { - G_2^J } ~~,~~c^{*}  = 1 + 2\alpha \,.
\ee

Moreover, the radial solution (\ref{eq.RadSolScattering}) in the near horizon of near extremal setup can be read as
\be\label{eq.RadSolNear}
R\left( {\bar \zeta} \right) = {\bar \zeta}^{\bar \alpha } \left( {1 - {\bar \zeta}} \right)^{\bar \beta } F\left( {\bar a^* ,\bar b^* ,\bar c^* ;{\bar \zeta}} \right)\,,
\ee
where
\be
{\bar \zeta} = \frac{\left( {\tilde r - \mu /2} \right)}{\left( {\tilde r + \mu /2} \right)}\,,
\ee
and the incorporating parameters are
\be
i\bar \alpha  = \sqrt {\bar G_1^J } \,,
\ee
\be
2\bar \beta  = 1 - \sqrt {1 - 4\bar G_3^J } \,,
\ee
\be
\bar a^*  = \bar \alpha  + \bar \beta  + i\sqrt { - \bar G_2^J } ~~,~~\bar b^*  = \bar \alpha  + \bar \beta  - i\sqrt { - \bar G_2^J }~~,~~ \bar c^*  = 1 + 2\bar \alpha \,,
\ee
and
\be
\bar G_1^J  = \frac{{\left( {\bar \omega  } \right)^2 }}{{\bar \lambda ^2 }}~~,~~ \bar G_2^J  =  - \left( {\frac{{\bar \omega }}{{\bar \lambda }} - \frac{{2r_c m\Omega _c }}{K}} \right)^2 ~~,~~ \bar G_3^J  =  - \frac{{\bar\lambda}}{K}\,.
\ee
Now let us consider the asymptotics of radial solution (\ref{eq.RadSolNear}) above. If one takes $\bar\zeta \gg \mu$, which corresponds to a large $\tilde r$, the radial solution above reduces to
\be
R(\bar\zeta)  \sim C_ +  {\bar\zeta}^{h - 1}  + C_ -  {\bar\zeta}^{ - h} \,.
\ee
In such asymptotics, one can get the corresponding retarded Green function in the form
\be
G_R  \sim \frac{{C_ -  }}{{C_ +  }}\,.
\ee
Explicitly in our near horizon of near extremal setup, the reading of this retarded Green function is

\be \label{GreenR}
G_R  \sim \frac{{\Gamma \left( {1 - 2h} \right)}}{{\Gamma \left( {2h - 1} \right)}}\frac{{\Gamma \left( {h - i\left( {{\bar G}_1^J  - {\bar G}_2^J } \right)} \right)\Gamma \left( {h - i\left( {{\bar G}_1^J  + {\bar G}_2^J } \right)} \right)}}{{\Gamma \left( {1 - h - i\left( {{\bar G}_1^J  - {\bar G}_2^J } \right)} \right)\Gamma \left( {1 - h - i\left( {{\bar G}_1^J  + {\bar G}_2^J } \right)} \right)}}\,.
\ee

Interestingly, the Green function above has an agreement with a two point function of the scalar operator in a CFT$_2$. For a scalar operator $\hat O$ that is dual to the scalar test probe, the two point function can be written as
\be \label{twopt1}
G\left( {{\tau} ^ +  ,{\tau} ^ -  } \right) = \left\langle {\hat O^\dag  \left( {{\tau} ^ +  ,{\tau} ^ -  } \right),\hat O\left( {0,0} \right)} \right\rangle \,,
\ee
where the left and right moving coordinates in a two dimensional worldsheet are given by ${\tau} ^ +$ and ${\tau} ^ -$, respectively. The two point function (\ref{twopt1}) is dictated by the conformal invariance and can be read as \cite{Chen:2010as,Chen:2010bh}
\be
G\left( {{\tau} ^ +  ,{\tau} ^ -  } \right) \sim \left( { - 1} \right)^{h_L  + h_R } \left( {\frac{{\pi T_L^J }}{{\sinh \left( {\pi T_L^J {\tau} ^ +  } \right)}}} \right)^{2h_L } \left( {\frac{{\pi T_R^J }}{{\sinh \left( {\pi T_R^J {\tau} ^ -  } \right)}}} \right)^{2h_R } e^{i\left( {q_L^J \mu _L^J {\tau} ^ +   + q_R^J \mu _R^J {\tau} ^ -  } \right)} \,,
\ee
with $\left(h_L,h_R\right)$ are the conformal dimensions, and $h_L = h_R =h$. It is understood that $\left(q_L^J,q_R^J\right)$ are the charges, $\left(T_L^J,T_R^J\right)$ are the temperatures, and $\left(\mu_L^J,\mu_R^J \right)$ are the chemical potentials that associate to the operator $\hat O$. In the equation above, we have incorporated the parameters in $J$-picture. The Euclidean correlator can be evaluated along the positive imaginary $\left\{\omega_{L}^J,\omega_{R}^J\right\}$ axis, i.e.
\be
G_E \left( {\omega _{L,E}^J ,\omega _{L,E}^J } \right) = G_R \left( {i\omega _{L,E}^J ,i\omega _{L,E}^J } \right)\,,
\ee
for a positive $\left\{\omega _{L,E}^J ,\omega _{L,E}^J\right\}$. In the theory with a finite temperature, $\omega _{L,E}^J $ and $\omega _{L,E}^J$ are discrete and take the values of Matsubara frequencies,
\be
\frac{{\omega _{L,E}^J }}{{2\pi T_L^J }} = m_L ~~{\rm and}~~\frac{{\omega _{R,E} }}{{2\pi T_R^J }} = m_R \,,
\ee
where $m_L$ and $m_R$ are integers.

Performing a Wick rotation to the 2D worldsheet coordinate, i.e.
\be
\tau^ +   \to i\tau _L ~~{\rm and}~~\tau^ -   \to i\tau _R \,,
\ee
where the Euclidean time $\tau_L$ and $\tau_R$ have the periodicity $2\pi/T_L^J$ and $2\pi/T_R^J$, respectively. Accordingly, the Euclidean correlator obtained from the two point function (\ref{twopt1}) can be written as
\[
G_E \left( {\omega _{L,E}^J ,\omega _{R,E}^J } \right) \sim {T_L^J}^{2h_L  - 1} {T_R^J}^{2h_R  - 1} \exp \left[ {i\left( {\frac{{\tilde \omega _{L,E}^J }}{{2T_L^J }} + \frac{{\tilde \omega _{R,E} }}{{2T_R^J }}} \right)} \right]\Gamma \left( {h_L  + \frac{{\tilde \omega _{L,E}^J }}{{2\pi T_L^J }}} \right)\Gamma \left( {h_L  - \frac{{\tilde \omega _{L,E}^J }}{{2\pi T_L^J }}} \right)
\]
\be \label{twopt2}
\times \Gamma \left( {h_R  + \frac{{\tilde \omega _{R,E} }}{{2\pi T_R^J }}} \right)\Gamma \left( {h_R  - \frac{{\tilde \omega _{R,E} }}{{2\pi T_R^J }}} \right)\,,
\ee
where
\be
\tilde \omega _{L,E}^J  = \omega _{L,E}^J  - iq_L^J \mu _L^J ~~{\rm and}~~\tilde \omega _{R,E}^J  = \omega _{R,E}^J  - iq_R^J \mu _R^J \,.
\ee
The agreement between (\ref{twopt2}) and (\ref{GreenR}) establishes the correspondence between a two dimensional conformal field theory and cosmological horizon in terms of scalar absorption calculation. It can be viewed as another supporting evidence for the Kerr/CFT correspondence of the cosmological horizon.

\section{Pair production}\label{sec.pair}

It has been shown that there exists a strong correlation between the existence of non-extremal Kerr/CFT correspondence and the pair production known as Schwinger effect near the horizon at near extremal state \cite{Chen:2012zn,Siahaan:2019ysk}. In the previous sections, we have shown that the Kerr/CFT holography dictionary can also be applied to the cosmological horizon. It motivates us to investigate the possibility for Schwinger effect to take place near the KNTNdS cosmological horizon. However, the required near horizon transformation is slightly different compared to the one in eq. (\ref{eq.transNearKerrCFT}) when we compute the central charge. The appropriate one to show the Schwinger effect is
\be \label{eq.nearhorizonpair}
r \to {\tilde r}_0  + \varepsilon \hat r ~~,~~dt \to \frac{{\left( {{\tilde r}_0^2  + a^2  + n^2 } \right)}}{{\Delta _0 \varepsilon }}d\hat t ~~,~~d\phi  \to d\hat \phi  + \frac{{\hat \Omega \left( {{\tilde r}_0^2  + a^2  + n^2 } \right)}}{{\Delta _0 \varepsilon }}d\hat t\,.
\ee
Note that we use the ``hat'' notation for the near horizon coordinate to distinguish it with the ``tilde'' one in (\ref{metric.nearJ}). In addition to the coordinate changes above, the mass is also shifted as
\be \label{eq.MassShift}
M \to M_0  + \frac{{\Delta _0 B^2 }}{{2{\tilde r}_0 }}\varepsilon ^2 \,.
\ee

Applying the transformations above to the spacetime metric (\ref{metricKNTNdS}) and the vector solution (\ref{eq.vector}) followed by taking $\varepsilon \to 0$ yields
\be\label{metricnearPair}
ds^2  =  - \frac{{\rho _0^2 }}{{\Delta _0 }}\left( {{\hat r}^2  - B^2 } \right)d\hat t^2  + \frac{{\rho _0^2 }}{{\Delta _0 \left( {{\hat r}^2  - B^2 } \right)}}d\hat r^2  + \frac{{\rho _0^2 }}{{\Delta _l  \Delta _x }}dx^2  + \frac{{Z^2 \Delta _l  \Delta _x }}{{\rho _0^2 }}\left( {d\hat \phi  + \frac{{2a{\tilde r}_0 \hat r}}{{Z\Delta _0 }}d\hat t} \right)
\ee
where $Z={\tilde r}_0^2 + a^2 + n^2$,
\be
\Delta _0  = 1 - \frac{{a^2  + 6\left( {{\tilde r}_0^2  + n^2 } \right)}}{{l^2 }} \,,
\ee
\be
\rho_0^2 = {\tilde r}_0^2 + (n+ax)^2 \,,
\ee
and the corresponding gauge field is
\be \label{vectornear}
A_\mu  dx^\mu   =  - \frac{{Q\left( {{\tilde r}_0^2  - \left( {n + ax} \right)^2 } \right)\hat r}}{{\rho _0^2 \Delta _0 }}d\hat t - \frac{{Q{\tilde r}_0 Z}}{{a\rho _0^2 }}d\hat \phi\,.
\ee
Interestingly, the result for near horizon geometry (\ref{metricnearPair}) looks similar to the near horizon metric (\ref{metric.nearJ}) that gives the central charge (\ref{centralchargeJ}). However, a closer look will show that these near horizon metrics are actually different from one another. It is understandable since the associated coordinate transformations to produce these near horizon metrics are slightly different. The coordinate transformation (\ref{eq.transNearKerrCFT}) incorporates the variables $r_c^*$ and $K$, which are basically the outcome of approaching $\Delta_r$ in KNTNdS metric (\ref{metricKNTNdS}) to be quadratic in $r$. Recall that such quadratic property is required to reveal the hidden conformal symmetry of the scalar probe. Accordingly, the associated central charge for the near horizon geometry (\ref{metric.nearJ}) is the one needed to produce the Cardy entropy (\ref{entropy}) after employing the left and right movers temperatures in the related hidden conformal symmetry generators. On the other hand, the near horizon geometry (\ref{metricnearPair}) and the corresponding coordinate transformation (\ref{eq.nearhorizonpair}) do not incorporate $r_c^*$ and $K$, as we consider the KNTNdS line element without any approximation for the $\Delta_r$ function. However, suppose one compute the central charge that correspond to the asymptotic symmetry of near horizon geometry (\ref{metricnearPair}), the resulting entropy after using Cardy formula and the suitable temperatures obtained in section \ref{s3.hidden} will not match the expected entropy for KNTNdS cosmological horizon. Nevertheless, for the case without cosmological constant, the near horizon of extremal black holes obtained by using the coordinate transformation (\ref{eq.nearhorizonpair}) can give the appropriate central charge to reproduce the related Bekenstein-Hawking entropy. For example, using the general central charge formula (\ref{c.formula}) for the near horizon of extremal Kerr-Newman black hole in \cite{Chen:2016caa} gives the correct central charge $c=12 Ma$  for the microscopic entropy calculation in $J$-picture \cite{Chen:2010ywa}. Note that ref. \cite{Chen:2016caa} discusses Schwinger effect near the Kerr-Newman black hole.

Now let us consider the Klein-Gordon equation for a charged massive test scalar in the near horizon of near extremal geometry constructed above. The general form reads
\be \label{eq.KGnear1}
\left( {\nabla _\mu   + iqA_\mu  } \right)\left( {\nabla ^\mu   + iqA^\mu  } \right)\Psi  = \mu^2 \Psi \,,
\ee
where $\mu$ is mass for the scalar field. To proceed, we employ the similar ansatz as the one we use to show the hidden conformal symmetry in section \ref{s3.hidden}. In the appropriate coordinate, it reads
\be
\hat \Phi  = \exp \left( { - i\hat \omega \hat t + i\hat m\hat \phi } \right)X\left( x \right)\hat R\left( {\hat r} \right) \,.
\ee
Subsequently, the resulting equations are
\be \label{eq.rad.pair}
\Delta _0 \partial _{\hat r} \left\{ {\left( {\hat r^2  - B^2 } \right)\partial _{\hat r} \hat R\left( {\hat r} \right)} \right\} + \left\{ {\frac{{\left( {Z\Delta _0 \hat \omega  - qQZ\hat r + 2{\tilde r}_0 a\hat m\hat r} \right)^2 }}{{Z^2 \Delta _0 \left( {\hat r^2  - B^2 } \right)}} - \mu ^2 Z - \lambda _l } \right\}R = 0 \,,
\ee
and
\be \label{eq.ang.pair}
\partial _x \left\{ {\Delta _l  \Delta _x X\left( x \right)} \right\} - \left\{ {\frac{{\left( {qQ{\tilde r}_0 Z - a\rho _0^2 \hat m} \right)^2 }}{{a^2 \Delta _x Z^2 \Delta _l  }} - \mu ^2 \left( {a\Delta _x  - 2nx} \right) - \lambda _l } \right\}X\left( x \right) = 0\,.
\ee

Accordingly, the effective mass in the radial equation (\ref{eq.rad.pair}) can be written as \cite{Chen:2012zn,Siahaan:2019ysk,Chen:2020mqs}
\be
m_{\rm eff}^2  = m^2  - \frac{{\left( {2{\tilde r}_0 a\hat m - qQZ} \right)^2 }}{{Z^3 \Delta _0 }} + \frac{{\lambda _l }}{Z}\,.
\ee
The corresponding Breitenlohner-Freedman bound that associates to the stability condition in the near horizon of AdS space, i.e. $m_{\rm eff}^2 \ge \tfrac{1}{4}L_{AdS}^{-2}$, can be expressed as
\be \label{eq.BF}
m_{{\rm{eff}}}^2  \ge  - \frac{{\Delta _0 }}{{4Z}}\,.
\ee
The violation of Breitenlohner-Freedman bound as stated in (\ref{eq.BF}) leads to the Schwinger effect near the cosmological horizon in KNTNdS spacetime. Similar to the pair production from black hole horizon in the near extremal state, such an effect can take place as well near the cosmological horizon. Indeed, the horizon must be in the near extremal condition, so it has a warped AdS structure that is needed in the prescription to show the Schwinger effect \cite{Chen:2012zn,Siahaan:2019ysk}.

\begin{figure}[h]
	\begin{center}
		\includegraphics[scale=0.5]{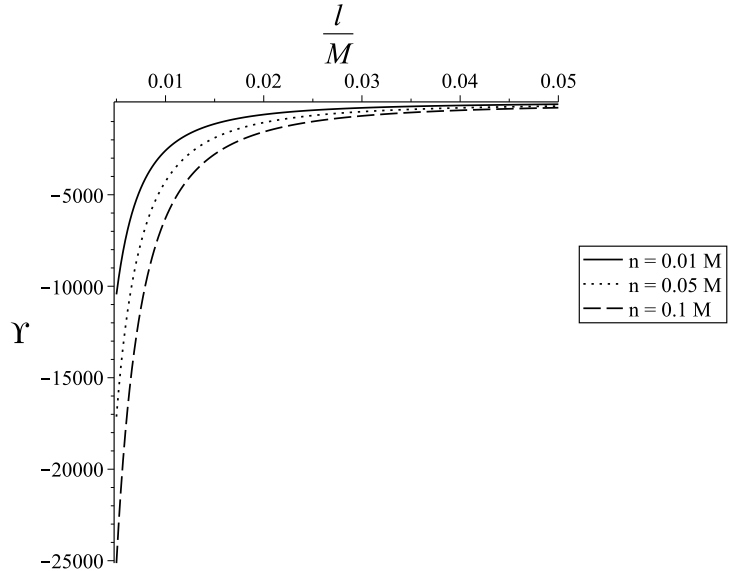}\caption{Plots for $\Upsilon =m_{{\rm{eff}}}^2  + \tfrac{{\Delta _0 }}{{4Z}}$}\label{fig.BF}
	\end{center}
\end{figure}

\begin{figure}[h]
	\begin{center}
		\includegraphics[scale=0.5]{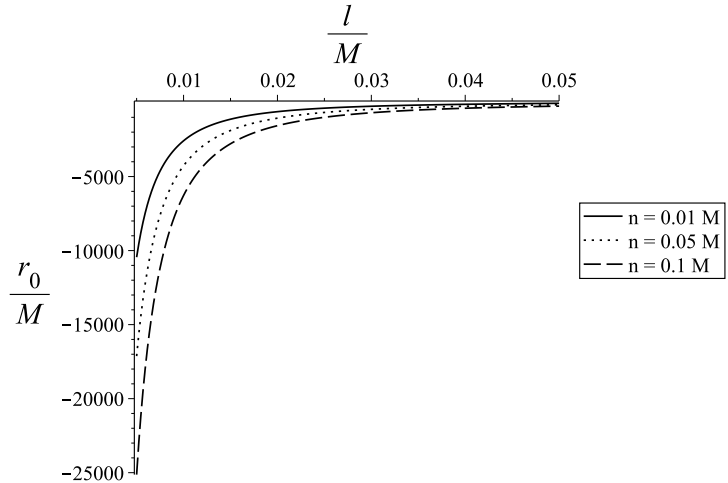}\caption{The values of ${\tilde r}_0$ that correspond to the plots in fig. \ref{fig.BF}.}\label{fig.r0}
	\end{center}
\end{figure}

We do not present here the thermal interpretation for the Schwinger effect above. But it can be done straightforwardly by following the prescription given in \cite{Chen:2012zn,Chen:2020mqs}. Illustrations for the Breitenlohner-Freedman bound violation are given in fig. \ref{fig.BF} where the corresponding ${\tilde r}_0$ values are evaluated in fig. \ref{fig.r0}. Note that we require ${\tilde r}_0$ to be real so the corresponding plots in fig. \ref{fig.BF} are well defined. The plots in fig. \ref{fig.BF} confirms the pair production can exist, and we can observe how NUT parameter affects the magnitude of $\Upsilon =m_{{\rm{eff}}}^2  + \tfrac{{\Delta _0 }}{{4Z}} <0$, where we can infer that the presence of NUT parameter yields the stronger violation of Breitenlohner-Freedman bound.

\section{Conclusion}\label{s6.conc}

In this work, we have presented the hidden conformal symmetry for the cosmological horizon in Kerr-Newman-Taub-NUT-de Sitter spacetime. As expected, the symmetry does exist for the cosmological horizon in a similar way for the black hole case. The method to construct the symmetry is straightforward by using the prescription that is normally employed for the black hole horizon. The obtained hidden conformal symmetry allows one to consider the Kerr/CFT holography for the cosmological horizon, such as in reproducing the entropy and the absorption cross section. The compatibility of Kerr/CFT holography for the cosmological horizon is expected due to the similarities between black hole and cosmological horizons, for example the associated entropy and Hawking radiation.

Interestingly, the $AdS_2 \times S^2$ or warped $AdS_3$ structure of the near horizon of a near extremal black hole can be used in showing the Schwinger effect near the black hole horizon. In fact, these structures of the horizon play an important role in computing the central charge in the Cardy formula. After showing compatibility of the Kerr/CFT holography for the cosmological horizon in KNTNdS spacetime, we show that the Schwinger effect can also occur near this horizon. Obviously such an effect can take place as well near the cosmological horizon of some special cases in the family of KNTNdS spacetime. That includes the \RN-NUT-de Sitter and the Kerr-NUT-de Sitter spacetime, including the vanishing NUT parameter consideration.

For the future works, we can investigate the hidden conformal symmetry and pair production near the cosmological horizon of dilatonic de Sitter \cite{Gao:2004tu}. Such study for the de Sitter geometry in low energy heterotic string theory \cite{Wu:2020cgf} can be attractive as well, considering the Kerr/CFT can be applied to this case \cite{Ghezelbash:2012qn}. Studying the pair production near the Kaluza-Klein black hole can also be interesting, due to its resemblances to the Kerr-Newman case.

\section*{Acknowledgement}
		
This work is supported by LPPM UNPAR under contract no. III/LPPM/2022-02/79-P. I thank the anonymous referee for his/her useful comments.

\end{document}